\documentclass[authoryear,nopreprintline]{elsarticle}
\usepackage[utf8]{inputenc}
\usepackage{graphicx} 

\usepackage[hidelinks]{hyperref}

\usepackage{amsmath,amssymb,amsfonts}
\usepackage{nicefrac}

\usepackage{sidecap}
\usepackage{wrapfig}

\title{Theories of synaptic memory consolidation and intelligent plasticity for continual learning}
\author[fmi,unibas]{Friedemann Zenke\corref{cor1}}
\author[fmi]{Axel Laborieux}

\address[fmi]{Friedrich Miescher Institute of Biomedical Research,  Basel, Switzerland}
\address[unibas]{Faculty of Science, University of Basel, Switzerland}
\cortext[cor1]{Correspondence: friedemann.zenke@fmi.ch}

\date{February 2024}

\usepackage[nohyperlinks]{acronym}
\acrodef{AI}[AI]{artificial intelligence}
\acrodef{ANN}[ANN]{artificial neural network}
\acrodef{MLP}[MLP]{multi-layer perceptron}
\acrodef{SI}[SI]{synaptic intelligence}
\acrodef{EWC}[EWC]{elastic weight consolidation}
\acrodef{IOF}[IOF]{ideal observer framework}
\acrodef{STC}[STC]{synaptic tagging and capture}
\acrodef{LTP}[LTP]{long-term potentiation}
\acrodef{LTD}[LTD]{long-term depression}
\acrodef{BNN}[BNN]{binarized neural network}
\acrodef{RL}[RL]{reinforcement learning}
\acrodef{iid}[i.i.d.]{independent and identically distributed}
\acrodef{BCM}[BCM]{Bienenstock-Cooper-Munro} 
\begin{document}


\maketitle

\section*{Abstract}
Humans and animals learn throughout life. Such continual learning is crucial for
intelligence. In this chapter, we examine the pivotal role plasticity mechanisms
with complex internal synaptic dynamics could play in enabling this ability in
neural networks. By surveying theoretical research, we highlight two fundamental
enablers for continual learning. First, synaptic plasticity mechanisms must
maintain and evolve an internal state over several behaviorally relevant
timescales. Second, plasticity algorithms must leverage the internal state to
intelligently regulate plasticity at individual synapses to facilitate the
seamless integration of new memories while avoiding detrimental interference
with existing ones. 
Our chapter covers successful applications of these principles to deep neural
networks and underscores the significance of synaptic metaplasticity in
sustaining continual learning capabilities. Finally, we outline avenues for
further research to understand the brain’s superb continual learning abilities
and harness similar mechanisms for artificial intelligence systems.

\section*{Key points}
\begin{itemize}
	\small
	\itemsep0em 
    \item The ability to continually learn, adapt to the environment, and retain memories of the past for extended periods is critical for intelligence.
    \item In this chapter, we survey a plethora of theoretical and modeling work suggesting that brains require plasticity mechanisms with two essential requirements to learn continually.
	\item First, these plasticity mechanisms have to evolve and maintain an internal state or memory over an extended range of timescales.
    \item Second, ``intelligent'' plasticity algorithms with access to this state regulate plasticity at individual synapses to form new memories whilst not causing harmful interference to the ones previously stored.
\end{itemize}

\tableofcontents
\clearpage

\section{Introduction}

A central building block of intelligence is memory. Memory lets us avoid costly
mistakes and adapt to changing environments quickly. Memory allows us to learn a
skill once and then repeatedly apply it in different situations. For instance,
if you remember the location of the nearest waterhole, you can sidestep a
tedious and potentially costly search every time you are thirsty. Similarly,
when we acquire a new skill, such as riding a bike, we do it once and not every
time we mount the saddle. Importantly, we can acquire new memories throughout
our lives, and we are also able to retain individual memories for a long time, 
sometimes for our entire life.

The vast neuronal networks that constitute our brain are the substrate
underlying this remarkable ability of memory formation and storage. Memories
are, in large part, stored as synaptic efficacy changes, i.e., the strength of the synaptic
connections between neurons. Jointly, these changes cause lasting changes to the
computational transformation that a given neuronal circuit performs, and it is
this change that lies at the heart of memory. For instance, consider a neuronal
circuit whose task is to associate a particular
reward, e.g., some food, with a specific sensory stimulus. Upon experiencing a
particular stimulus, e.g., a tone of a given frequency, the circuit in question may
not produce any output initially. That is to say that it has not yet associated
any value with this stimulus. Only once the tone is paired consistently with a
reward does the brain circuit start producing an output specific to the paired
stimulus. Thus, the next time the circuit receives the tone stimulus, its output
will signal the value, e.g., the expected reward, even before the food has
arrived. Meanwhile, the circuit response to another control stimulus, i.e.,
another tone, is unchanged --- the circuit described above stores a memory of the
rewarded stimulus. The paradigm described above is called Pavlovian
conditioning, which honors Pavlov, who first performed such an experiment on his
dog. And importantly, there are several brain areas, like, for instance, the
amygdala, that are involved in such value learning.

\begin{SCfigure}[10][tbh]
    \centering
    \includegraphics[width=0.5\textwidth]{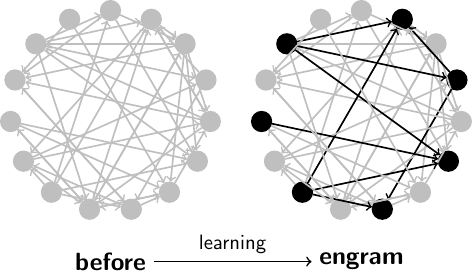}
    \caption{Schematic of engram formation through learning. The sum of changes written into a network to form a memory is called the engram (black). }
    \label{fig:engram}
\end{SCfigure}

However, let us stay abstract and continue considering our hypothetical brain
circuit.  To achieve this change of input and output, our neuronal circuit in
question has to make alterations to its internal connections to accomplish the
input-specific change of its output.  The sum of all of these changes to the
connections is also called the ``engram'' (Fig.~\ref{fig:engram}). This term
refers to the physical changes to the network architecture that are written into
the network to store the memory, i.e., the associated reward or value in the
above example. Thus, memory storage is intricately linked to altering synaptic
connections in neuronal circuits in an experience-dependent and orchestrated
way.  But how do individual synapses know how to change their connections to
accomplish changes to the output of an entire circuit? This is one of the great
open questions of circuit neuroscience.

Many bio-chemical mechanisms
exist that can change synaptic connections in an activity-dependent manner.
These mechanisms are jointly called synaptic plasticity and are presumed to be
the bio-physical basis of memory and learning. This book covers synaptic
plasticity in-depth, so we won’t delve into them too deeply. For now, let us
focus on the conceptual challenges these synaptic plasticity mechanisms must overcome to
implement memory. There are several challenges. First, the synaptic changes must
be stimulus-specific to form a distinct memory. Second, the changes need to be
orchestrated so that the circuit does not succumb to pathological activity but
remains in a healthy activity regime. Thus, plasticity also needs to ensure the
stability of the circuit. This function is typically associated with various
forms of homeostatic plasticity in neuronal circuits that act on different
length- and timescales. Finally, plasticity must ensure that the circuit
remains plastic for storing future memories without compromising the information
already stored in the circuit.

This last bit is crucial as it is often vital for us to store more than one
memory. There are many examples, but an everyday instance for humans is
language. Each spoken word is a specific sequence of sounds. We can associate
meaning with each word if we know the language. For language, we know from our
own experience that we acquire it gradually. We first learn simple words such as
``mama'' and ``papa.'' Later, our vocabulary grows. At some point, we may even learn
another language. These learned associations have to be stored somehow in
synaptic changes in some network in our brain. Because we acquire new pieces of
information one at a time, we must incrementally add new memories to the same
network without erasing everything previously learned.
                                            
Our ability to learn continually is pervasive and intrinsically linked to how we
learn. It applies to most learning problems requiring memory. Language is only
one example. Other examples are physical skills such as riding a bicycle or
playing volleyball. 
We must learn both skills in an often painstaking process, and they require overlapping sets of muscles, so at least some of the neuronal circuits involved are bound to be shared among them.
We learn both tasks at different times, and yet
learning to play volleyball does not impair our ability to ride a bike. 
A plethora of mechanisms enable us to learn continually.
Some of them act at the system level, i.e., neuronal
circuits or networks with different learning dynamics \citep{mcclelland_why_1995, schapiro_complementary_2017, van_de_ven_brain-inspired_2020}, while
others operate at the synaptic level. 
System level and replay approaches, also widely used in machine learning, are covered elsewhere in this book.
This chapter focuses on the putative synaptic mechanisms.

In the following, we first review the classic Hopfield network, an
influential associative memory model, which allows us to define essential
concepts such as memory capacity and interference.
The latter is linked to forgetting, and we will see how.
Equipped with these concepts, we will explore several strands of research suggesting that synapses must possess an internal state or memory in their own right and intelligently use it to store new memories. 
To understand why, we will get to know the notion of memory lifetime
and the fundamental requirement for memory consolidation mechanisms in the
continual learning setting. 
Finally, we will turn to the more contemporary deep
\ac{ANN} models, the problem of catastrophic forgetting,
and putative solutions based on synaptic consolidation and ``intelligent''
synaptic plasticity rules.

\section{Learning and forgetting in Hopfield networks}

When studying theoretical models of learning and memory, there are few models as iconic as the Hopfield model \citep{amari_learning_1972, hopfield_neural_1982} which constitutes a simple and elegant model of associative memory.
In the Hopfield model, we consider a recurrent network model consisting of $n$ units or neurons with all-to-all connectivity given by the following connectivity matrix 
\begin{equation}
W_{ij} = \frac{1}{n} \sum_{\mu=1}^p \xi_i^\mu \xi_j^\mu
\label{eq:hopfield_learning_rule}
\end{equation}
which stores $p$ randomly and independently drawn binary patterns $\xi \in \{-1,1\}^n$ in which a one appears with a probability of 0.5.
By convention neuron $i$ is active in memory pattern $\mu$ if $\xi_i^\mu=+1$.
If we look at the above expression, we see that each term of the sum adds positive connections between neurons that are active (+1) in the same memory together.
Conversely, if one neuron is active (+1) and the other inactive (-1) in a given pattern, the sign of the weight update is negative resulting in an inhibitory interaction between the two.
Hence, we can think of learning in the Hopfield model as a Hebbian learning rule which defines assemblies of neurons with mutually excitatory connections as engrams for each stored memory (\citealp{hebb_organization_1949}; cf.\ Fig.~\ref{fig:engram}).

\begin{SCfigure}[10][tb]
\centering
\includegraphics[width=0.6\textwidth]{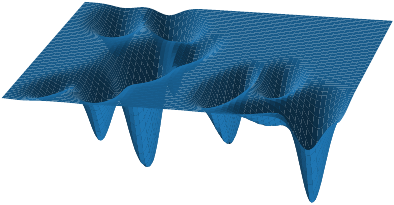}
\label{fig:hopfield_energy}
\caption{Schematic of an energy landscape in the Hopfield model. 
The stored memories correspond to minima in the energy landscape. When the network is cued with a partial memory as an input, it is initialized near one of the potential wells. The network dynamics then evolve to minimize the energy, thereby recalling the previously stored memory.}
\end{SCfigure}

Now once the memories are stored in the connections of the model, how do we retrieve the stored memories?
In the Hopfield model this is done through the network dynamics.
The dynamics of the neuronal state $\vec s$ is such that in every time step $t$ one neuron is updated at a time such that
\begin{equation}
   s_i(t) = \mathrm{sign}\left( \sum_j W_{ij} s_j(t-1) \right) \quad .
   \label{eq:hopfield_dynamics} 
\end{equation}
This update strategy is also known as the asynchronous update. 
When applying this update over and over again one can show that the dynamics follow an energy function and the stored patterns $\xi^\mu$ are local minima of this energy landscape (Fig.~\ref{fig:hopfield_energy}) such that their corresponding neuronal state $\vec s$ is a stable point attractor, i.e., the state does not change any more.

Why is this a good model of memory? 
Well, suppose we initialize the neuronal state vector close to one of the stored patterns. 
Without loss of generality let us assume we initialize the state $\tilde s_i = \xi_i^1 + \mathrm{noise}$ where the noise is binary noise that randomly flips elements of the vector to the other state. Now $\tilde s_i$ is a noisy version of $\xi_i^1$. 
If the noise is not too strong and the original pattern is still discernible, the Hopfield network's dynamics are such that it gradually restores the stored pattern.
This property constitutes memory recall in the model. 
In other words, we query the network by showing a slightly different input to it and the internal dynamics converge to one of its stored memories.
The dynamics stop once the pattern is restored.

That this happens is easy to see if we only store a single pattern. 
In this case the connectivity matrix is given by $W_{ij} \sim \xi^1_i \xi^1_j$.
Now let us assume that $\vec s(t=0)=\vec \xi^1$, then the update equation is the following:
$$ s_i(t) = \mathrm{sign} \left( \frac{1}{n} \sum_j \xi^1_i \xi^1_j \xi_j^1 \right) = \mathrm{sign} \left( \xi^1_i \right) = s_i(t-1) $$
because $\sum_j \xi^1_j \xi_j^1 = n$.

In short, learning corresponds to storing the memories using the ``learning rule'' given by Eq.~\eqref{eq:hopfield_learning_rule} and memory recall is achieved by letting the network dynamics converge to one of the minima in the energy landscape.

\subsection{Memory capacity and forgetting}

Although not initially formulated in a continual learning setting, we can
appreciate that forming new memories adds additional terms to the sum in
Eq.~\eqref{eq:hopfield_learning_rule}.  The natural question, then, is about the
network's memory capacity, i.e., how many patterns can we store in a given
network?  Perhaps not surprisingly, the answer is that the memory capacity
$\alpha=\frac{p_\mathrm{max}}{n}$ is finite.  Assuming uncorrelated patterns,
one can compute the maximum capacity using replica methods from statistical
physics \citep{amit_storing_1985, hertz_introduction_1991}.  The answer is
$\alpha \approx 0.138$.

What is more surprising is what happens when the network exceeds this capacity.
When reaching capacity, the Hopfield network undergoes a phase transition at
which it completely loses its ability to recall any stored
memory.  The theoretical origin of this transition is the same as a
ferromagnet's transition when heated up to its Curie temperature, upon which it
loses its ferromagnetic properties.

In the Hopfield model, the phenomenon is called ``blackout catastrophe,'' but it is also often referred to as ``catastrophic forgetting'' in the literature.
Here, a word of caution is in order, as it is the first time we encounter
the term catastrophic forgetting, and we will re-encounter it when discussing
learning in \acp{ANN} (Section~\ref{sec:learning_in_anns}).  We highlight it
here because the term refers to fundamentally different phenomena in these two
contexts.  In the Hopfield model, catastrophic forgetting is linked to its
finite memory capacity. It describes a phenomenon that impairs the network's
ability to recall any stored memory patterns through its  dynamics.  In
the context of \acp{ANN}, as we will see later, it refers to old memories being
quickly overwritten by new memories, a feature also called the palimpsest
property.  What adds to the confusion is that overwriting old memories through
new memories is the solution to avoiding catastrophic forgetting in the
Hopfield model. 
At the same time, forgetting is the primary problem for \acp{ANN}.
While the whole meaning of this statement will only become clear at the end of
this chapter, let us first understand why overwriting old memories solves the issue of
catastrophic forgetting in the Hopfield network. 

Because catastrophic forgetting
occurs at the capacity limit, a simple remedy is ensuring the network never
reaches this critical capacity. Formally, this can be achieved by dropping the
early terms in the sum in Eq.~\eqref{eq:hopfield_learning_rule} or more
plausibly by adding a small weight decay that shrinks the contribution of older
patterns that ensures forgetting of old memories at a rate such that the
critical capacity is never reached \citep{nadal_networks_1986}. Finally, the
problem of overloading can also be avoided by preventing unbounded growth of the
synaptic weights.
That is why gradual forgetting is a necessary and healthy consequence of the finite capacity of the Hopfield model.
Hence, avoiding catastrophic forgetting requires a strategy to gracefully forget old memories to ``make room'' for new ones. 
Thus, learning and forgetting are fundamentally tied together and 
neuronal circuits are immediately faced with a dilemma.

\subsection{The stability-plasticity dilemma}

Neuronal circuits have to weave new memories into the same synaptic fabric that holds older memories while ensuring that old memories are not forgotten too quickly since this precludes the system from operating close to its capacity. 
This fundamental dilemma affects many, if not all, memory systems with a finite capacity, and it is often called the stability-plasticity dilemma or the palimpsest paradox
\citep{nadal_networks_1986, grossberg_competitive_1987, amit_learning_1994, abraham_memory_2005}.

It is called a paradox because the above statement contradicts our daily
experience by which we constantly and quickly form new memories. 
For instance, we readily remember a person to whom we are introduced. 
But we also retain old memories,
some of them for life. Our brains have evolved to deal with this
plasticity-stability dilemma very efficiently. They learn continually by adding
new memories with minimal impact on many old memories. The underlying synaptic and neuronal
circuit mechanisms accomplishing this feat are poorly understood. The general
idea is that our brains use different active gating mechanisms for memory
storage and memory systems whose dynamics evolve on vastly different timescales
to consolidate memories for long-term storage. 
While some of these consolidation
mechanisms presumably act at the system level \citep{mcclelland_why_1995, schapiro_complementary_2017}, 
for instance, by actively replaying previously experienced activity patterns \citep{roxin_efficient_2013, van_de_ven_brain-inspired_2020, tome_coordinated_2022},
in this chapter, we focus on the putative synaptic
mechanisms, typically referred to as ``synaptic memory consolidation.''

\section{Synaptic memory consolidation and synaptic complexity}

Synaptic memory consolidation has been studied experimentally and theoretically, and the term can refer to a broad spectrum of effects or models. 
On the one hand, there are phenomenological synaptic plasticity models capturing experiments associated with the ideas of synaptic tagging and capture. 
On the other hand, normative models seek to fathom the fundamental limits of memory lifetimes. 
Finally, contemporary work started to appreciate the need for synaptic consolidation in deep neural networks. 
While we still lack a unified theoretical framework for synaptic memory consolidation, all theories share the fundamental concept of ``synaptic complexity.''
What do we mean by that? 

In classic neural network models, the state of synapses is commonly characterized by a single scalar quantity denoting its efficacy or weight. This starkly contrasts synapses in neurobiology, which boast intricate internal biochemical signaling. 
Thus, rather than single scalar quantities, synapses are complex dynamical systems in their own right. 
Synaptic memory consolidation theories assume that this added complexity accounts for metaplasticity and underlies memory consolidation. 
In this section, we cover the core ideas that introduced synaptic complexity in the models before turning toward \acp{ANN} in the next section.

\subsection{Phenomenological models of metaplasticity}

In neurobiology, many intracellular signaling molecules are known to be involved in expressing synaptic plasticity and triggering lasting changes to the rules of plasticity themselves. 
The diverse activity-dependent mechanisms involved in the plasticity of the plasticity rules are called ``metaplasticity'' \citep{abraham_metaplasticity:_2008}. 
Classically, one of the putative functional roles associated with metaplasticity is the stabilization of otherwise unstable Hebbian plasticity, an essential requirement for storing any memory in a given memory system \citep{zenke_hebbian_2017}. 
For instance, the classic \ac{BCM} model posits that each neuron maintains an estimate of its average postsynaptic firing rate to implement a metaplastic form of firing rate homeostasis \citep{bienenstock_theory_1982}. 
Substantial experimental efforts were subsequently put into finding such a moving threshold with some success \citep{cooper_bcm_2012}. 
However, these efforts also uncovered a host of synaptic forms of metaplasticity that do not point at a neuron-wide threshold \citep{abraham_metaplasticity:_2008}. 
Much of this reported synaptic complexity and metaplasticity have since inspired new models and theories of memory consolidation and continual learning in \acp{ANN} \citep{jedlicka_contributions_2022}. 
We will talk more about metaplasticity later.

\subsection{Phenomenological models of synaptic tagging and capture}

If metaplasticity was one of the reasons for adding synaptic complexity to the models, the \ac{STC} hypothesis was the other main driver. The hypothesis is concerned with the putative cellular and synaptic mechanisms leading to stabilization processes that enable synapses to retain their efficacy for a prolonged duration of time despite the continuous replacement of their constituent proteins \citep{redondo_making_2011}.
The \ac{STC} hypothesis is mainly concerned with memory encoding and at which stage in the process the decision is made whether memory should be stabilized or not. It is, thus, a hypothesis about the persistence of memory encoding. It suggests that while a memory event may induce \ac{LTP} or \ac{LTD} at a given synapse, whether this change is persistent or fades away is determined by a set of interacting mechanisms that can be triggered at a specific time. The initial and transient memory encoding is called “tagging” because synapses are tagged with the tentative information. In contrast, their long-term stabilizing requires capturing plasticity-related proteins due to a trigger event.

Studying the implications of the \ac{STC} hypothesis in a theoretical context required models
that captured the tagging experiments and the rich phenomenology of plasticity induction observed in vitro.
This requirement motivated adding an internal state and dynamics to the synaptic models.
The first models that captured the hypothesis' core ideas used a state evolving in a double-well potential for modeling synaptic tags and consolidation variables \citep{clopath_tag-trigger-consolidation:_2008}.
Thus, the model assumed internal synaptic dynamics following some energy function, effectively changed by plasticity induction protocols and plasticity-related proteins. A similar contemporary model relied on discrete states with probabilistic transitions  \citep{barrett_state_2009}.
Conceptually, it is clear that the phenomenology of the presumed synaptic tag and the subsequent consolidation calls for added synaptic complexity. In fact, capturing the wealth of synaptic tagging and capture dynamics involves more than one hidden state and its associated dynamics  \citep{ziegler_synaptic_2015}.

While the role of some of the experimentally observed effects remains elusive, the utility in stabilizing memory is obvious. So, it is not surprising that it was demonstrated in modeling studies. For example, \citet{luboeinski_memory_2021} showed that a calcium-based synaptic plasticity model encodes memories through strongly interconnected cell assemblies. A later memory recall was substantially improved when supplemented with an \ac{STC} mechanism.

\begin{figure}[tbp]
    \centering
    \includegraphics[width=1.0\textwidth]{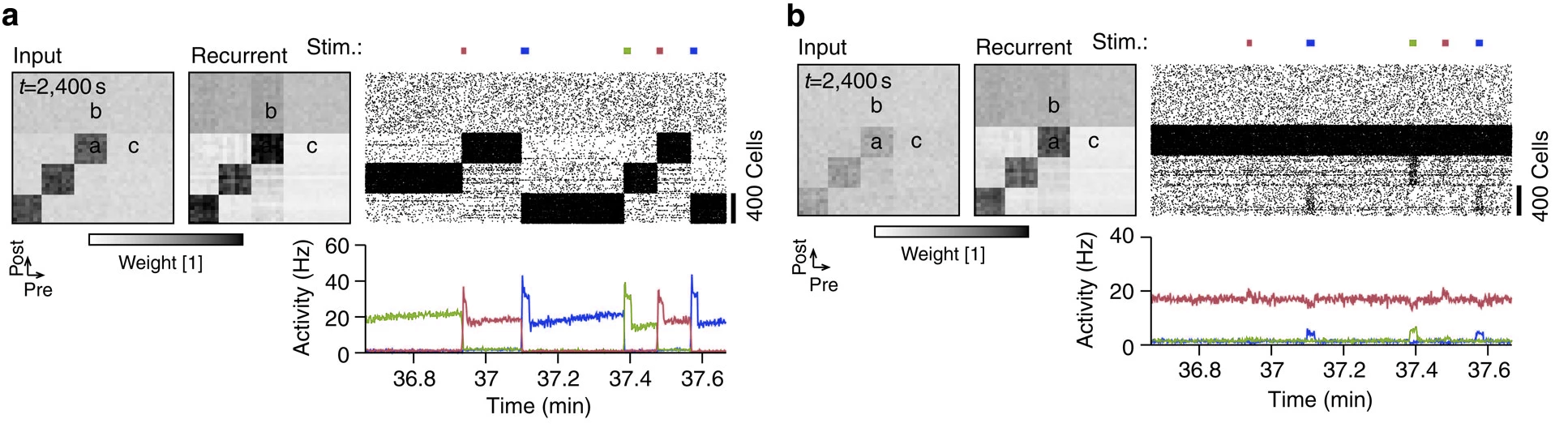}
    \caption{\textbf{(a)}~Snapshot of network connectivity and activity during associative learning and recall in a plastic spiking neural network model with synaptic consolidation \citep{zenke_diverse_2015}. 
    Subsections of the input and recurrent weight matrix corresponding to excitatory connections after reordering the neurons according to assembly membership (left). Assemblies are visible as blocks on the diagonal.
    Spike raster of neuronal activity of the corresponding neurons (right).
    Stimulus input is indicated by colored bars (top).
    Population firing rate of neurons with specific tuning to one of the three stimuli over time (bottom).
    After repeated exposure to a set of three external stimuli for 2400s, neurons in the plastic spiking neural network simulation have formed stimulus selective ensembles with strong recurrent connectivity which allows them to maintain elevated firing rates during the delay interval after the stimulus has been removed.
    \textbf{(b)}~Same as panel (a), but without synaptic consolidation.
    The afferent synaptic weights have largely lost their stimulus selectivity and the recurrent network dynamics decouple from the external world.
    Despite ongoing external stimulation (top), only the third assembly remains active. Figure reproduced from \citep{zenke_diverse_2015}.}
    \label{fig:consolidation_snn}
\end{figure}

Similarly, another study focusing on memory storage and recall in spiking neural networks found that synaptic consolidation was crucial for recalling stored memories in a persistent working memory-like state \citep{zenke_diverse_2015}. The model drew heavily on the notion of associative memory. But instead of directly manipulating neurons as is commonly done in the Hopfield model, the authors also learned the feedforward afferent synapses to the recurrent network, a reasonable assumption for an associative memory model in the brain.
Perhaps counter-intuitively, the authors found that consolidation was of primary importance on the afferent connections.
The reason was that once a set of memory assemblies had formed in the recurrent part of the network and could be recalled through selective delay activity (Fig.~\ref{fig:consolidation_snn}a), synapses within the network would primarily see recurrent reverberating activity that did not coincide with stimulus-driven activity in the afferent connections.
This lack of input activity caused afferent synapses to slowly depotentiate or forget, thereby, leaving the formed assemblies to decouple from the external world (Fig.~\ref{fig:consolidation_snn}b) lest they were equipped with synaptic consolidation dynamics (cf.\ Fig.~\ref{fig:consolidation_snn}a). 
In a later study \citet{tome_coordinated_2022} extended the above study to multiple distributed brain regions and found that repeated memory reactivation can explain system consolidation through coordinated synaptic plasticity, provided the timescales for synaptic consolidation are chosen sensibly.

The above findings give a glimpse of the improved memory performance of plastic neural networks equipped with synaptic consolidation mechanisms, which were mainly introduced to account for experimental observations.
And capturing the rich phenomenology of plasticity induction and synaptic tagging requires models with complex synaptic dynamics. 
However, more theoretically rooted considerations also call for synaptic complexity.

\subsection{Synaptic consolidation and fundamental limits on memory lifetimes}

If a memory system's capacity is finite and creating new memories requires overwriting old memories, how can memory storage be orchestrated efficiently in such systems? When addressing this question from a normative angle, it is helpful to consider the memory lifetime, i.e., the time a given memory can be read out successfully from a memory system. 
To gain analytic insight into memory lifetimes, working in a general framework that makes a minimal set of assumptions about the memory system is convenient. 
In our case, we are interested in storing memory traces in synapses through synaptic plasticity, and a suitable framework to study memory lifetimes is the \ac{IOF}.

\subsubsection{The Ideal Observer Framework}

The \ac{IOF} simplifies the learning problem by focusing at individual synapses while ignoring all the complexity of the surrounding network and even the learning algorithm.
Although a stark simplification, the \ac{IOF} makes the learning problem theoretically more tractable and allows deriving fundamental insights about the effect of synaptic complexity on learning. 
Specifically, we assume that our memory system acquires new memories with the constant rate $r$. 
A constant rate of memory acquisition immediately links the memory capacity and the lifetime of a memory. 
Further, we assume that storing each memory is associated with synaptic changes that either 
depress or potentiate each synapse or leave it unchanged. 
In reality these plasticity events are generated by some learning algorithm, about which we only make few assumptions.
We mainly require that the plasticity events are balanced, i.e., the up and down events cancel each other on average. 
This assumption is reasonable. Without it all synapses would soon saturate at their respective maximum or minimum, e.g., with an exponential distribution \citep{fusi_limits_2007}.
Finally, we assume that each event is associated with storing a memory and that these memories are statistically independent.

\begin{SCfigure}[10][tbh]
    \centering
	\includegraphics[scale=0.8]{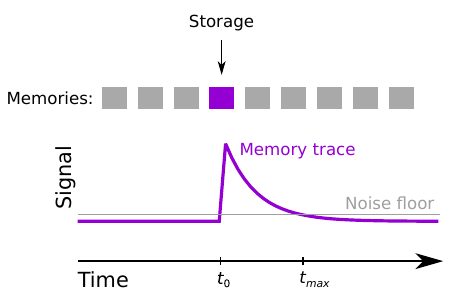}
	\caption{Schematic of the lifetime of a memory in the \ac{IOF}. 
	Memories depicted by the gray blocks are stored at a constant rate $r$.
	We pick an arbitrary memory as our reference and record its memory trace ($t_0$).
    At storage time the associated signal is maximal.
	However, due to interference caused by storing other memories, the memory trace of our
	reference decays over time.
	Once it drops below the noise floor at $t_\mathrm{max}$ even an ideal observer cannot recall the memory any more.}
	\label{fig:iof}
\end{SCfigure}

Without loss of generality, we now pick any memory as our reference and ask under which conditions an ideal observer, i.e., a hypothetical observer that has noise-free unimpeded access to all synaptic states could still tell this reference memory, the ``signal,'' from the noise caused by storing other memories (Fig.~\ref{fig:iof}).
Thus,  we ask how many memories we can store before the memory trace of this reference is lost in noise.
Because we make no assumptions about the network and the mechanisms related to recalling the memory, the \ac{IOF} instead allows us to focus on the best case scenario. 
In other words, any memory capacity derived within this framework is naturally an upper bound and any practical implementation in a physical system would perform worse.

\subsubsection{Limits on memory lifetimes in simple synapses}
\label{sec:life_time}

Let us first consider a memory system composed of $N_\mathrm{syn}$ simple binary synapses.
Storing a memory in our system corresponds to flipping the state of some of them with probability $q$. 
The initial ``signal'' of this memory is proportional to $\sim q N_\mathrm{syn}$.
We are now interested in how this signal degrades over time in comparison to the fluctuations in synaptic efficacy due to ongoing plasticity with the rate $r$.
Since storing every additional memory causes a synapse to flip with the constant  probability $q$, the waiting time of each synapse to randomly flip is exponentially distributed. 
As a direct consequence, the signal of the reference memory also fades exponentially $\sim e^{-qrt}$ (cf.\ Fig.~\ref{fig:iof}).
These considerations allow computing the expected memory lifetime 
\begin{equation}
    t_\mathrm{max} \sim \frac{\ln\left(q \sqrt{N_\mathrm{syn}}\right)}{qr} \quad ,
    \label{eq:memory_lifetime_simple_synapse}
\end{equation}
a troubling result due to its logarithmic dependence \citep{amit_learning_1994}.
Synapses are ideal loci of information storage in neuronal circuits because of their sheer abundance.
However, the logarithmic dependence in Eq.~\eqref{eq:memory_lifetime_simple_synapse} largely nullifies this advantage. 
The only choice we are left with for extending memory lifetimes is to set $q \ll 1$. 
Still, a small storage probability $q$ is also undesirable because it reduces our initial signal.
Formally the above theoretical considerations underscore the fundamental character of the plasticity-stability dilemma.
Further, the result shows that not even an idealized observer can leverage the sheer abundance of synapses ($N_\mathrm{syn} \gg 1$) to sidestep the problem.
The above considerations raise the important question of how biological systems can store many memories with a large initial signal to noise ratio and safeguard them over extended periods of time.

\subsubsection{Memory lifetimes in complex synapses}
\label{sec:complex_synapses}
\begin{figure}[tbh]
    \centering
    \includegraphics[width=\linewidth]{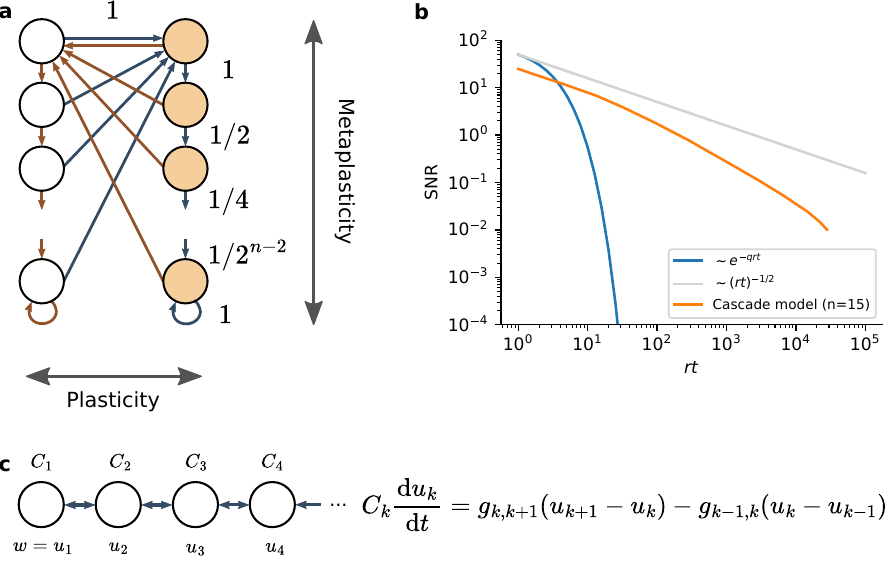}
    \caption{Power-law forgetting in the Cascade model.
    \textbf{(a)}~Schematic of the Cascade model. Synapses are either in a weak (left column, white circles) or strong (right column, orange circles) efficacy state, but have an internal metaplastic state represented vertically. The transitions between states are governed by probabilities that decrease exponentially with the depth in the cascade. 
    \textbf{(b)}~Example forgetting curves, which show the signal to noise ratio as a function of the number of time with storage rate $r$, for an exponential decay (blue), a Cascade model with cascade depth $n=15$ (orange; data from \citealp{fusi_cascade_2005}), and a power-law with exponent $- \nicefrac{1}{2}$ (gray) aligned to the initial SNR of the exponential case.
    Experimentally observed forgetting curves typically follow a power-law with exponents smaller than one.
    \textbf{(c)}~The \citet{benna_computational_2016} model for continuous-valued synapses. The full state of the synapse is a chain of variables $u_1$, $u_2$, ... interacting through differential equations reminiscent of connected beakers of areas $C_k$ and pipe widths $g_{k, k+1}$. The efficacy of the synapse is the first variable $u_1$. 
    All panels were redrawn from \citet{fusi_cascade_2005} and \citet{benna_computational_2016} respectively.}
    \label{fig:cascade_model}
\end{figure}

\citet{fusi_cascade_2005} proposed an elegant solution to this problem which led to the formulation of the Cascade model, one of the first models to leverage synaptic complexity for extended memory storage.
The idea underlying the Cascade model is the following:
Instead of two synaptic states, corresponding to high and low efficacy, each synapse possesses a cascade of states for each of the two efficacy levels (Fig.~\ref{fig:cascade_model}a). 
When a synapse is potentiated and receives another potentiating event, it is pushed down further in the corresponding cascade.
While this does not alter the strength of the synapse, it alters its probability for future weight changes in such a way that it becomes increasingly difficult to depress a synapse that has progressed far in the cascade. 
In other words, the deeper a synapse is in the cascade, the less likely it becomes for its efficacy to change.
The position in the cascade of a synapse thus constitutes a form of metaplasticity \citep{abraham_metaplasticity:_2008}.

As simple as this mechanisms may appear, \citet{fusi_cascade_2005} showed that such added synaptic complexity has profound consequences for the memory lifetime. Instead of exponentially fading memory traces, the Cascade model approximates power-law forgetting, i.e., much longer memory lifetimes (Fig.~\ref{fig:cascade_model}b).

Since the Cascade model makes specific assumptions about the basic structure of the allowed transitions between the discrete cascading states, it is not clear whether such transition probabilities are optimal for memory storage.
\citet{lahiri_memory_2013} addressed this question by analyzing all possible models while allowing all-to-all transition probabilities between the states.
They derived a theoretically attainable upper bound on the achievable memory lifetime, and found that the bound was not tight for existing models.
Thus, that there is still room for improvement.

While the discrete state structure may be seen as a limitation of the model, \citet{benna_computational_2016} later generalized the synaptic model to a system of interacting continuous variables with different interaction strengths (Fig.~\ref{fig:cascade_model}c), and showed that this could lead to similarly increased memory lifetime in which the SNR decreases as $\sim \nicefrac{1}{\sqrt{t}}$ while also ensuring almost linear capacity scaling in the number of synapses, a significant improvement over previous models with $\sim \nicefrac{1}{t}$ forgetting, once more cementing
the fundamental relevance of synaptic complexity for memory storage in neural networks.

The \ac{IOF} outlined above demonstrates the potential of synaptic complexity and consolidation to mitigate the plasticity-stability dilemma. 
However, to enjoy power-law forgetting in an actual memory system requires working with real-world data, which violates some of the simplifying assumptions made above. 
It also requires a recall mechanism which may have to content itself with less than ideal observations of the system state.
While promising studies with a concrete recall mechanism exist that show the above ideas generalize to real-world data \citep{ji-an_face_2023}, modeling such systems remains challenging because it requires dealing with high-dimensional data and a functional network model operating on these data, the latter of which is a major feat in itself.
Here, deep learning has emerged as a practical tool for neuroscience that sidesteps the issue by providing a recipe for building complex neural networks using optimization principles \citep{richards_deep_2019}.

\section{Continual learning in artificial neural networks}
\label{sec:learning_in_anns}

\acp{ANN} underlie the majority of modern \ac{AI} applications.
Typical applications include playing the board games Chess or Go at the grandmaster level, generating images from a text prompt, or the powerful language models capable of passing the Turing test.  
With their biological counterparts, these networks share the fundamental property that 
neuronal activity has to propagate through multiple synaptic processing layers before the network generates an output.
Computer scientists refer to the property of having many processing layers as ``depth,'' hence the term deep neural networks \citep{goodfellow_deep_2017}.
As in biological brains, deep neural networks store memories in their weights.
Storing information in \acp{ANN} requires a learning algorithm that changes the connection strengths to minimize an error at the network's output.
Curiously, the sheer size of modern deep learning models and the associated cost of training them, along with the increasing need to deploy them in mobile applications and autonomous agents, is creating widespread demand for effective continual learning algorithms that allow adding new experiences to these systems as they go and without needing to retrain the entire system. 

Because \acp{ANN} are functional neural networks that operate on real-world data and due to their commonalities with biological neural networks in neurobiology, \acp{ANN} have emerged as an indispensable tool for studying memory and continual learning in neuroscience \citep{richards_deep_2019}.
Interestingly, standard training algorithms in deep learning, unlike the brain, require a stationary data distribution, i.e., a fixed training set from which one samples randomly and uniformly during training.
When confronted with non-stationary data, the standard learning setting for humans and animals, deep neural networks succumb to catastrophic forgetting \citep{parisi_continual_2019, kudithipudi_biological_2022}.

\subsection{Catastrophic forgetting}

What is catastrophic forgetting in \acp{ANN}? 
The concept is perhaps best illustrated through an example.
Suppose you want to train a multi-layer network to classify images of handwritten digits.
The procedure to do so is basically the following.
First, you set up a network in which the initial synaptic weights are randomly initialized.
Due to this randomness the network computes a random function and in all likelihood not the one that happens to reliably classify handwritten digits.
Acquiring this ability requires training the network.
However, since we want the network to learn continually, we split this problem into several tasks.
Task~A is to classify zeros and ones.
Task~B requires distinguishing twos and threes and so forth.

\begin{figure}[tb]
    \centering
    \includegraphics[width=1.0\textwidth]{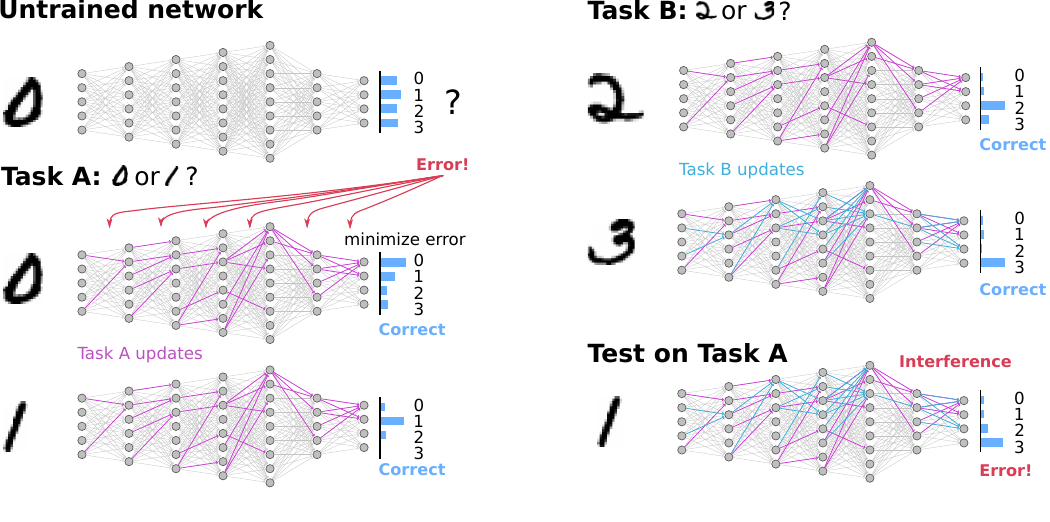}
    \caption{Illustration of catastrophic forgetting in \acp{ANN}. 
    The initially untrained network (top left) is first trained successfully to classify 0 and 1 digits, denoted Task~A, which results in task-specific weight updates in purple (bottom left). 
    Then, the network is trained on a second task to classify 2 and 3 digits, denoted Task~B, which results in new task-specific weight updates in blue (top right). 
    After training on Task~B, the performance on Task~A is degraded due to interference between the memory traces associated with both tasks (bottom right).}
    \label{fig:learning_algo_deep_nets}
\end{figure}

To train the network on Task~A one datum is presented to the network (Fig.~\ref{fig:learning_algo_deep_nets}).  
Let us assume we show it a pixel image of the digit ``zero,'' but due to the lack of training, the network's output says ``one.''
Thus, the network made an error.
The goal of the training algorithm is to reduce this error, by attributing each weight in the network by how much it has contributed to producing the erroneous output. 
Then the weights are updated iteratively such that the error is reduced.
This step is called credit assignment and in most cases an algorithm called ``backpropagation of error'' is used to compute these weight updates \citep{goodfellow_deep_2017}.
After processing many data in this way the network eventually learns to correctly classify ones and zeros, while generalizing to examples that it has not seen during training.
The network has learned to master Task~A.

Now suppose we want to learn to also classify between ``twos'' and ``threes.''
This corresponds to Task~B.
We can repeat the above procedure, but in all likelihood we will update some connection weights that were essential for solving Task~A, the Zero-One classification problem, thereby affecting the previously learned solution adversely (Fig~\ref{fig:learning_algo_deep_nets}).
Thus, as we improve classification accuracy on Task~B, we are forgetting about the previous task.

\begin{figure}[tb]
    \centering
    \includegraphics[width=1.0\textwidth]{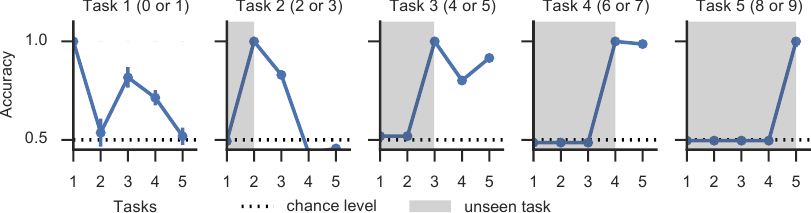}
    \caption{Catastrophic forgetting in an \ac{ANN} trained on split-MNIST. The plots show the task accuracy for the different tasks on the Split MNIST problem as a function of the number of tasks the network has seen during training. 
    Split MNIST consists of five consecutive tasks. Task~1 corresponds to classifying zeros and ones from MNIST, Task~2: Twos and Threes, and so forth. 
    Since each task is a binary classification problem, the network performs at chance level on unseen tasks (gray shaded region). 
    After training on Task~1, its accuracy is close to~1. Similarly Task~2 accuracy rises close to 100\% after training on Task~2. 
    However, meanwhile Task~1 accuracy drops close to chance level.
    A similar pattern repeats for the other tasks.
    The error bars correspond to SEM ($n=10$).
    Figure adapted from \citep{zenke_continual_2017}.}
    \label{fig:split_MNIST_catastrophic_forgetting}
\end{figure}

The example of learning to classify pairs of handwritten digits is a simple benchmark called split-MNIST \citep{zenke_continual_2017}.
This simple benchmark nicely illustrates the gravity of the problem of catastrophic forgetting in \acp{ANN} (Fig.~\ref{fig:split_MNIST_catastrophic_forgetting}).
It also highlights that unlike in the Hopfield network discussed above, in \acp{ANN} catastrophic forgetting is not due to a capacity limit.
We know that the same \ac{ANN} has no problem classifying all ten digits as long as they are trained jointly.
Here, training jointly means that the network is repeatedly shown random samples of all classes, i.e., all ten digits.
Thus the problem of continual learning in \acp{ANN} is linked to the learning algorithm itself and how it stores memories in the network.

The struggle of \acp{ANN} to learn continually contrasts with our everyday experience of how we learn. We and animals tend to learn better when different tasks are learned independently. Moreover, we are capable of lifelong learning. 
Several circuit mechanisms have been proposed that may underlie this unique ability of biological brains \citep{parisi_continual_2019, kudithipudi_biological_2022}. 
First, the idea of complementary learning systems posits that different brain areas learn on separate timescales and make continual learning possible through their interactions \citep{mcclelland_why_1995, schapiro_complementary_2017}. The key idea is that one area, e.g., the hippocampus, acquires new memories quickly. 
While also susceptible to fast forgetting, it then copies this memory to a different brain area, e.g., the neocortex which learns and forgets more slowly. 
This avenue is also called system-level consolidation and may require some form of replay \citep{roxin_efficient_2013, van_de_ven_brain-inspired_2020,  tome_coordinated_2022}. It is covered elsewhere in this book. 
Second, synaptic consolidation requires synaptic complexity and is thought to accomplish a similar feat within a given neuronal circuit or brain area. This idea requires synapses and neurons to keep track of the information they store and then use this knowledge to inform synaptic plasticity to intelligently weave new memories into the neural memory fabric to minimize interference with previously stored information. We have already discussed phenomenological and theoretically motivated models that argued for synaptic complexity. 
Specifically, we looked at metaplasticity as prescribed by the Cascade model and how it can extend memory lifetimes. 
Thus far, the empirical success of applying the Cascade model to deep neural networks was limited, possibly because several of its underlying assumptions are violated by deep learning algorithms and real-world data. 
For instance, weight updates are not balanced, and the stored memories are not uncorrelated. Historically, synaptic theories of continual learning applied to \acp{ANN} thus took the idea of synaptic complexity but employed it differently as we will see next. 
We will cover the central ideas underlying recent continual learning algorithms that mitigate catastrophic forgetting in deep neural networks.
After that, we will highlight two recent studies that have reinvigorated aspects of the original ideas of the Cascade model. 
But before discussing any of these topics, we must understand how \acp{ANN} learn from data by optimizing loss functions.

\subsection{Learning from data as an optimization problem}

Unlike the models above, the learning rules for \acp{ANN} are not defined explicitly, but the learning problem is usually formulated as an optimization problem, commonly referred to as training. To train a deep neural network, one starts from a blank slate, a network architecture in which the initial synaptic weights are drawn randomly and independently. This randomly initialized network typically does not do anything useful. To train the network, one defines a scalar loss function that measures how well the network solves a particular task that one would like it to solve. In supervised learning, one of the most commonly used training approaches, one typically measures the distance of the network output to many predefined target labels that one collects in a large training dataset. The lower the loss, the better the network does on the training data. The goal of the optimization algorithm behind the training is to adjust the parameters or weights in the \ac{ANN} such that the loss function assumes a local maximum or minimum. 

\begin{figure}[tbh]
    \centering
    \includegraphics[width=0.5\textwidth]{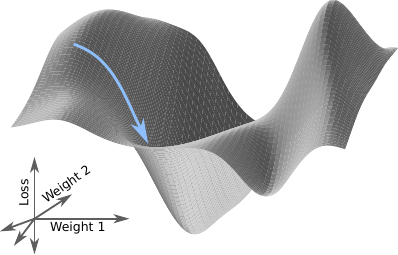}
    \caption{Schematic of an ANN optimization landscape. The optimization landscape is defined by the value of the scalar loss function and spanned by the network parameters, i.e., the synaptic weights. 
    Due to the vast number of synapse in \acp{ANN} the optimization landscape landscape is high-dimensional. Learning in an \acp{ANN} corresponds to finding minima in this high-dimensional landscape. 
    Here we can only visualize two synapses. 
    Typically this optimization is achieved by following the negative gradient (blue line).}
    \label{fig:optimization_landscape}
\end{figure}

The loss function is a useful theoretical abstraction for picturing the learning problem in \acp{ANN}.  We can think of this loss function as a landscape spanned by the parameters, e.g., the synaptic weights of the network (Fig.~\ref{fig:optimization_landscape}). It is a high-dimensional landscape with as many dimensions as there are parameters in the network. While these parameters include synaptic weights and other parameters like firing thresholds, we typically speak of only synaptic weights for brevity in the following section. Weight updates are then easily obtained within this framework by following the landscape's negative gradient, which can be computed efficiently using the back-propagation of the error algorithm. The scalar loss function takes its global minimum when the network's outputs closely resemble the desired outputs, or ``labels.'' 
Since gradient descent is only guaranteed to find a local minimum of the loss landscape, one worry is that the optimization could get stuck at a local minimum. However, years of deep learning research suggest that local minima are rarely a problem when one follows a few practical tricks of the trade, which includes using a sensible weight initialization strategy and certain architectural assumptions like convolutional layers or attention mechanisms \citep{goodfellow_deep_2017}. Gradient descent has proven so successful in deep learning that it is the go-to method for neural network training.

But how do the above concepts of memory formation and the notion of continual learning correspond to such an optimization landscape view?
To answer this question, let us first introduce some notation that will be useful later. We will refer to the loss function of task $\mu$ as $\mathcal{L}_\theta^\mu = \mathcal{L}(\theta,x_\mu,y_\mu)$, where $x_\mu$ and $y_\mu$ are the inputs and labels belonging to task $\mu$
and $\theta$ are the network parameters, including the synaptic weights. 

\begin{figure}[tb]
    \centering
    \includegraphics[width=\textwidth]{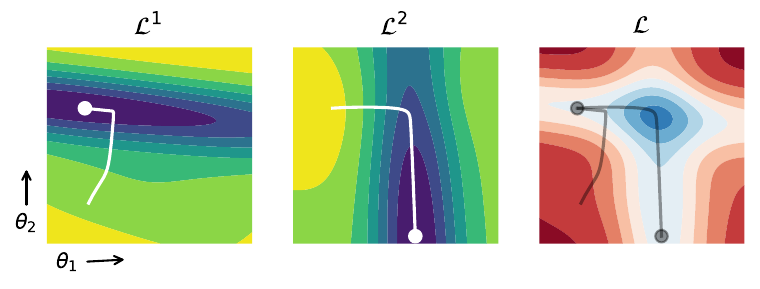}
    \caption{Catastrophic forgetting viewed through the lens of optimization theory. 
    Heat maps of the loss landscape in two parameter directions $\Theta_{1/2}$  (darker colors correspond to lower loss).
    Learning to solve Task~1 corresponds to minimizing $\mathcal{L}^1$ (left).
    An example learning trajectory following the negative gradient is shown in white.
    The final parameters are marked by the point.
    Subsequent learning of Task~2 corresponds to  
    minimizing $\mathcal{L}^2$ (middle). 
    This minimization inadvertently makes parameter changes that increase $\mathcal{L}^1$ which results in gradual forgetting of Task~1. 
    As a result, the parameters following the optimization are far away from the global minimum of the combined loss $\mathcal{L}$ of both tasks (right).
    }
    \label{fig:optimization_landscape_tasks}
\end{figure}

Forming a memory in a deep network corresponds to moving from a high loss region to one with a low loss value. Thus, learning is akin to finding a valley or minimum in the landscape (Fig.~\ref{fig:optimization_landscape_tasks}). 
That is, when the network learns Task~1, this corresponds to a change of its parameters $\theta$ that minimize $\mathcal{L}^1$. 
We will refer to $\theta(\mu)$ as the network's parameters after learning task $\mu$. Similarly, we can think of forgetting as moving to a region with high loss.

What happens when we want to learn another task in the continual learning setting? As mentioned above, the loss depends on the data and thus the optimization landscape changes over time.
In other words, every task has a different optimization landscape defined on top of the same coordinate system spanned by the synaptic weights. 

From our above example, we can immediately see how learning new tasks can cause problems for previous tasks. 
For instance, we may inadvertently increase the task loss $\mathcal{L}^1$ by minimizing the loss $\mathcal{L}^2$ (Fig.~\ref{fig:optimization_landscape_tasks}).
Thus, learning on Task~2 causes us to forget Task~1.
In practice, this forgetting can be severe and not just a minor effect, which is apparent from the Split-MNIST task (cf.\ Fig.~\ref{fig:split_MNIST_catastrophic_forgetting}).
The goal of any continual learning algorithm is to find parameters $\theta^*$ that perform well on all tasks. 
In our optimization landscape view, we want to minimize 
the global objective given by the sum of the task losses
\begin{equation}
    \mathcal{L}_\theta = \sum_\mu \mathcal{L}_\theta^\mu \quad .
    \label{eq:overall_objective_function}
\end{equation}

However, the learning algorithm only has access to the inputs and labels of the current task. 
To solve this feat, it is key that not all weights are equally important for learning a specific task (cf.\ Fig.~\ref{fig:learning_algo_deep_nets}).
This asymmetry allows an intelligent learning algorithm to minimize changes to recently stored memory engrams to prevent them from being overwritten. 
To that end, the learning algorithm must protect those synapses that contributed most to learning previous tasks and had the strongest influence on decreasing their corresponding loss values.

To implement such a learning algorithm, 
we need two ingredients.
First, we need some way of telling how important every synapse is. 
We call this the synaptic importance measure. 
Second, we need a mechanism to protect the essential synapses from updates that would degrade the memory they store. 
We first discuss a protection mechanism that directly relates to synaptic consolidation.

\subsection{Alleviating catastrophic forgetting through synaptic consolidation}

Let us look at a concrete protection mechanism originally introduced by \citet{kirkpatrick_overcoming_2017}.
It is again best understood by example.
Let us consider the familiar scenario in which we want to learn two tasks.
First we train on Task~1 by minimizing the corresponding loss $\mathcal{L}^1$.
But when training on Task~2 we instead minimize the following surrogate loss:
\begin{equation}
\mathcal{\tilde L}_\theta^2 = \mathcal{L}_\theta^2 + \sum_i \frac{\lambda}{2}\Omega_i(\theta_i - \theta_i(1))^2 \quad ,
\label{eq:ewc_loss}
\end{equation}
where we have added a second term to the normal task loss.
It is a sum that runs over all weights $\theta_i$ and quadratically penalizes deviations from their previous weight value $\theta_i(1)$ after learning Task~1. 
Moreover, each term in the sum is weighted by the synaptic importance value $\Omega_i$ and a global strength parameter $\lambda$.
The added term is reminiscent of parameter regularization commonly used in deep learning and we can see 
that it penalizes large changes to parameters that were good at solving Task~1.
Because the penalty term's quadratic potential is the same as the energy stored in an elastic spring, \citet{kirkpatrick_overcoming_2017} called this protection mechanism ``elastic'' weight consolidation.

If the penalty incurred for changing a weight was the same for all weights, then it would do no good. 
Penalizing any weight change would simply prevent us from learning Task~2. 
So we need to make sure to only penalize those weight changes that we do not want because they would destroy previously stored memories.
This is where the notion of synaptic importance comes in.
It quantifies how important a given weight is for storing previous memories.
Remember, one of our core assumptions is that some synapses are more crucial for storing previous memories or tasks than others.
In Eq.~\eqref{eq:ewc_loss}
important weights $i$ have a large $\Omega_i$ value compared to less important weights. 
This formulation biases the learning algorithm to preferentially change weights with smaller importance values.

\begin{figure}[tb]
    \centering
    \includegraphics[width=1.0\textwidth]{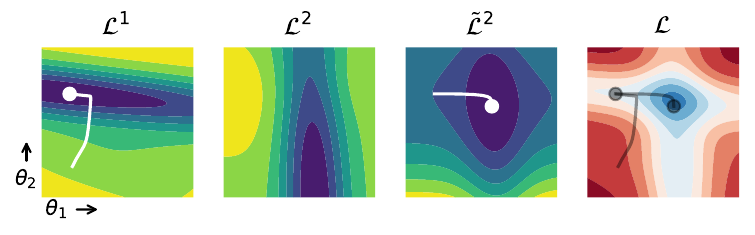}
    \caption{Avoiding catastrophic forgetting with synaptic consolidation. 
    Heat maps of the loss landscape in two parameter directions $\Theta_{1/2}$  (darker colors correspond to lower loss) for the same tasks as shown in Fig.~\ref{fig:optimization_landscape_tasks}.
    The third panel shows the surrogate loss $\mathcal{\tilde L}^2$ in which a quadratic penalty has been added (cf.\ Eq.~\ref{eq:ewc_loss}).
    Optimizing the surrogate loss yields final parameters $\Theta$ close the minimum of the global loss $\mathcal{L}$ (rightmost panel).}
    \label{fig:optimization_landscape_tasks_SI}
\end{figure}

In an ideal scenario, the quadratic penalty closely approximates the shape of the loss landscape $\mathcal{L}_1$. 
In this case gradient descent on the surrogate loss $\mathcal{\tilde L}^2$ takes us close to the actual minimum of $\mathcal{L}$ at which both tasks are solved (Fig~\ref{fig:optimization_landscape_tasks_SI}).

The question is how to find suitable $\Omega_i$ values.
How can individual synapses ``know'' how important they are for storing existing memories in the network? 
There are multiple ways of defining synaptic importance, which
relate to the geometry of the optimization landscape.

\subsection{Synaptic importance measures}

In \ac{EWC}, \citet{kirkpatrick_overcoming_2017} use the diagonal of the Fisher information metric as the importance measure $\Omega_i$.
The Fisher information metric measures how much information a random variable carries about a parameter of interest. In the context of neural networks, it captures how sensitive the model's predictions are to changes in the parameters. 
Close to a minimum the Fisher information metric is equivalent the Hessian of the loss.
The Hessian is a square matrix of second-order partial derivatives that describes the curvature of the loss function. 
Thus \ac{EWC}'s importance measure characterizes the local curvature of the loss landscape. 
That is, it assigns a high value to highly curved parameter space directions. 
Suppose we just finished training on Task~1. By definition, we are close to a minimum of $\mathcal{L}^1$ in parameter space. 
As we learn Task~2 next, we want to minimize parameter changes that change $\mathcal{L}^1$ because then we would make things worse on Task~1, i.e., we would forget. 
Thus, we want to make sure that we minimize our loss on Task~2 while ideally only walking into flat directions on $\mathcal{L}^1$. Flat means less curvature. 
To accomplish this, \ac{EWC} more strongly penalizes updates in directions with high curvature on previous tasks. 
In practice, this strategy works remarkably well and has since become a standard for comparison.

However, \ac{EWC} requires the Fisher information metric's diagonal, which must be computed in a separate phase. 
Moreover, the Fisher metric is a point estimate at the minimum and thus lacks information about the global loss landscape.  
Finally, it is difficult to conceive how actual synapses in the brain could compute the Fisher information metric.

\subsection{Estimating synaptic importance online}

To address the above points, \citet{zenke_continual_2017} proposed \ac{SI} with an importance measure that can be computed locally at the synapse and online during learning while retaining information of the global loss landscape over the entire optimization trajectory.
The algorithm rests on a mathematical identity that ties together each synapses contribution to decreasing the loss. 
Specifically, let $\omega_k$ be the contribution of synapse $k$ to decreasing the loss during training:
\begin{equation}
    \mathcal{L}^\mu-\mathcal{L}^{\mu-1} = - \sum_k \omega_k = \sum_k \int_{t^{\mu-1}}^{t^\mu} g(\theta(t))_k  \theta_k'(t) dt 
\end{equation}
where  $g=\nabla_\theta \mathcal{L}$ is the gradient on the current task, $\theta'$ is the parameter update, and the integral runs from the beginning of training on Task~$\mu$ to the end.
For gradient descent one typically chooses $\theta' = -\eta g$ with some positive learning rate $\eta$. If we further drop the sum, the expression simplifies to:
\begin{equation}
    \omega_k = \eta \int_{t^{\mu-1}}^{t^\mu} g(\theta(t))_k^2 dt \quad .
\end{equation}

That means if we assume a gradient-following learning algorithm, each synapse can estimate its importance for a given task by simply computing the sum of all squared weight updates it encounters during training. 
Importantly, it can do this locally because the weight updates, i.e., the potentiation and depression events, are known to the synapse. 
However, doing so requires synaptic complexity. 
In other words, the synapse must possess an internal state and internal dynamics to perform this type of bookkeeping. 

\begin{figure}[tb]
    \centering
    \includegraphics[width=1.0\textwidth]{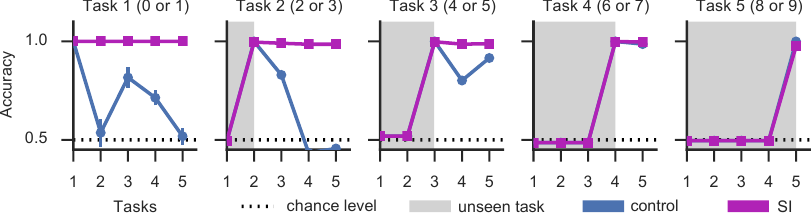}
    \caption{Synaptic intelligence prevents catastrophic forgetting on Split-MNIST. 
    Same as Figure~\ref{fig:split_MNIST_catastrophic_forgetting}, but with added datapoints for \ac{SI}.
    Error bars correspond to SEM ($n=10$).
    Figure adapted from \citep{zenke_continual_2017}.}
    \label{fig:split_MNIST_SI}
\end{figure}

By further combining the above synaptic quantity with the synaptic change during training on Task~$\mu$ yields the required synaptic importance measure (cf.\ Eq~\eqref{eq:ewc_loss}):
\begin{equation}
    \Omega_k^\mu = \frac{\omega_k^\mu}{\Delta_k^{2}(\mu) + \epsilon} 
\end{equation}
with $\Delta(\mu) \equiv \theta(\mu)-\theta(\mu-1)$ and some small positive parameter $\epsilon$ to prevent division by zero. 
In practice, the idea of protecting important synapses with the above surrogate loss has profound impact on the learning dynamics and can drastically reduce forgetting in deep neural networks (Fig.~\ref{fig:split_MNIST_SI}) comparable to \ac{EWC}.

Notably, the work outlined above shows that synaptic consolidation of important synapses overcomes catastrophic forgetting in deep neural networks. 
Crucially, with \ac{SI}, we know a practical and straightforward algorithm capable of estimating synaptic importance locally, a prerequisite for implementing similar algorithmic principles in the biological wetware.

\subsection{Task-agnostic continual learning with Cascade models}

The methods introduced above rely on a synaptic importance measure associated with each parameter, which affects their propensity to undergo future plasticity.
In practice this is accomplished through an auxiliary loss added to the optimization objective. 
However, this notion has a critical drawback as it requires discrete task boundaries and explicit knowledge of when a given task ends as it requires updating the auxiliary loss. 
In other words, the learning algorithms are task-aware as synapses must be consolidated after learning a specific task and before learning another. 
A direct consequence of this requirement is that regularization-based approaches such as \ac{SI} consolidate synaptic changes on the timescale of a single discrete task. 
This fundamentally differs from the idea behind the Cascade models introduced in the previous section. 
Cascade models do not require explicit knowledge of the task boundaries, and their synaptic consolidation dynamics span a wide range of timescales across several orders of magnitude. 
In the following we see that variations of this idea can be applied to deep \ac{RL} and \acfp{BNN}.

\subsubsection{Reinforcement learning with complex synapses}

\Acf{RL} is a paradigm where an agent evolves through the states $s_t$ of an environment and make decisions on which action $a_t$ to take at any given time.
The goal of the agent is to maximize the expected future rewards, as given by a scalar objective function $(s_t , a_t) \mapsto r(s_t, a_t)$ defined over the states and actions.
To do this, the agent needs to maintain a replay buffer, a set of observations $\{s_t, a_t, r(s_t, a_t)\}$ it encountered during its exploration of the environment, and use it for training its reward model.
Learning is done by adding new observations to the buffer, and shuffling it to ensure that the data is \ac{iid}.
The \ac{iid} requirement can be interpreted as a way for the agent not to forget past experiences when learning the most recent ones, which can be understood as within-task forgetting.
Therefore, continual learning can potentially improve \ac{RL} by alleviating the need for the replay buffer within a single task.
However, regularization-based methods are not well-suited for this objective as they typically require knowledge about task transitions.
On the other hand, metaplasticity models such as \citet{benna_computational_2016} are promising candidates.

\begin{figure}[tb]
    \centering
    \includegraphics[scale=1.0]{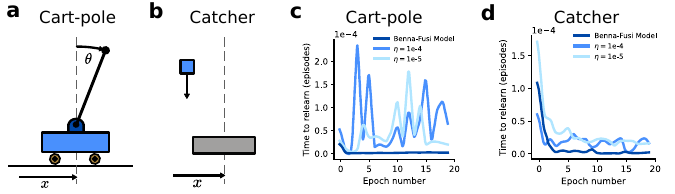}
    \caption{Reinforcement learning with complex synapses. 
    \textbf{(a)}~Schematic of the Cart-pole task. 
    The task consists of a cart with a rod on a pivot. The task is to apply forces to the cart, which can move along the $x$-direction, such that the pole remains upright ($\theta=0$). 
    \textbf{(b)}~Schematic of the Catcher task. 
    The task is to move the gray pallet along the $x$ direction to catch the falling cubes.
    \textbf{(c)}~Time to relearn Cart-pole after learning Catcher as a function of number of epochs for complex synapses defined by the Benna-Fusi model and for vanilla gradient descent for different learning rates $\eta$.
    Complex synapses clearly reduce the time for relearning on this task. 
    \textbf{(d)}~Same as (c), but for the Catcher task after learning Cart-pole.
	Data in panels (c) and (d) extracted from \citet{kaplanis_continual_2018}.}
    \label{fig:rl}
\end{figure}

\citet{kaplanis_continual_2018} investigated the benefits of using the metaplasticity model of \citet{benna_computational_2016} in the context of an \ac{RL} problem.
To that end, they trained a deep \ac{RL} agent, which uses an \ac{ANN} to generate actions from sensory stimuli, to learn two different tasks alternately.
Specifically, they trained the agent on the Cart-Pole and Catcher tasks (Fig.~\ref{fig:rl}a,b) and recorded the time needed to relearn the previous task to quantify how well the agent remembered (Fig.~\ref{fig:rl}c,d).
They found that the agent equipped with metaplastic complex synapses was quicker at relearning the previous task, while the control agent with conventional synapses was unable to relearn Cart-Pole after learning Catcher (Fig.~\ref{fig:rl}c).
The authors interpreted this phenomenon as an extreme case of catastrophic interference.
These findings suggests that metaplastic synapses can reduce forgetting when learning multiple tasks in \acp{ANN}.

In another set of experiments, \citet{kaplanis_continual_2018} investigated whether complex synapses could prevent within-task forgetting by alleviating the need for the replay dataset usually required to train deep \ac{RL} agents.
Considering the same tasks as above, the authors found for the Cart-Pole task that only the agent equipped with complex synapses could successfully learn online without the replay dataset, whereas vanilla agents were unable able to learn across a wide range of different learning rates.
In contrast, the vanilla agents were able to learn 
the Catcher task without replay, which the authors explained by the more stationary experience distribution of the task.
Overall, the authors concluded that synaptic metaplasticity can effectively reduce forgetting in \acp{ANN} without requiring an task-aware consolidation mechanism as in \ac{EWC}.
Still, it remains an open question how scalable this approach is and whether it transfers to more difficult deep \ac{RL} problems like those studied by \citep{kirkpatrick_overcoming_2017}.

\subsubsection{Synaptic metaplasticity in binarized neural networks}

\begin{figure}[tbp]
    \centering
    \includegraphics[width=1.0\linewidth]{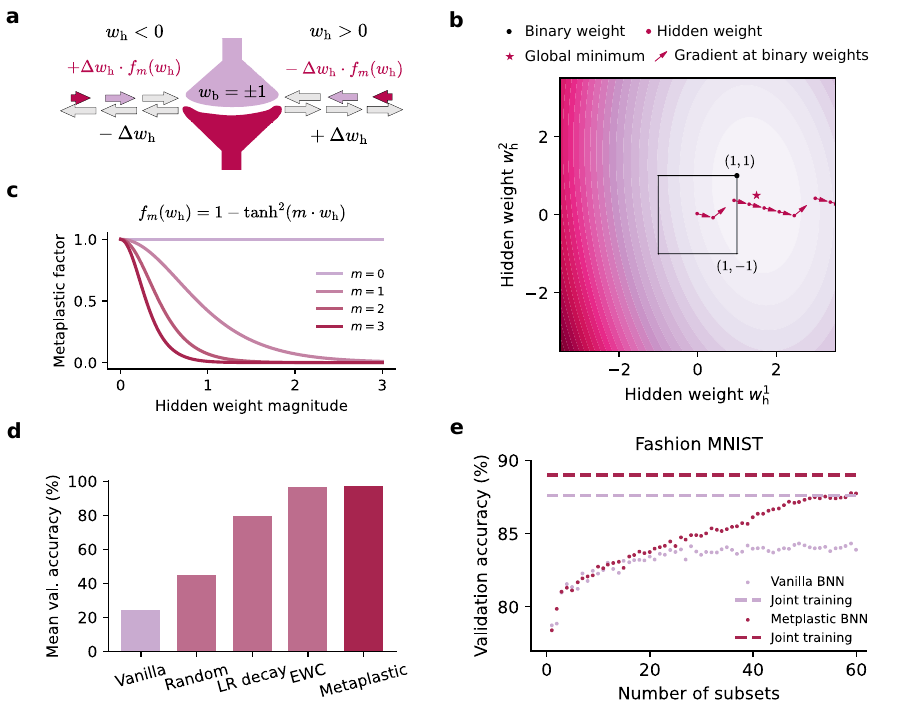}
    \caption{Complex synapses improve continual learning in \acp{BNN}. 
    \textbf{(a)}~Schematic of a metaplastic binary synapse. 
    The binary synapse can take one of two efficacy values: $w_\textrm{b}=\mathrm{sign}(w_\textrm{h})=\pm 1$, as a function of the underlying real-numbered state variable $w_\textrm{h}$.
    Gradient-based weight updates are directly applied to $w_\textrm{h}$ during learning if they have the same sign as $w_\textrm{h}$ (bottom arrows), otherwise the updates are multiplied by a multiplicative factor $f_m(w_\textrm{h})$, which is a monotonically decreasing function of $|w_\textrm{h}|$ (top arrows).
    \textbf{(b)}~Weight evolution during training in the phase plane of a two-dimensional binarized model optimized on a quadratic objective function Eq.~\eqref{eq:bnn_training}.
    The box corresponds to the possible binarized weight values.
    The global minimum is not achievable with binarized weights.
    As a result, the hidden weights are consistently updated with the gradient at binarized weights $(1,1)$ or $(1,-1)$.
    $w_{\rm h}^1$ is the most important weight and grows in magnitude, the less important weight $w_{\rm h}^2$ stays close to $0$.
    \textbf{(c)}~Plot of the decreasing multiplicative factor $f_m(w_\textrm{h})$ applied to weight updates that decrease with the norm of $w_\textrm{h}$.
    \textbf{(d)}~Mean validation accuracy after training sequentially on six permuted versions of the MNIST dataset for different training algorithms.
    Vanilla: no consolidation, Random: consolidation of a random set of weights, LR decay: decreasing learning rates for each new task, EWC: \acl{EWC}, and with metaplastic \ac{BNN}.
    Metaplastic synapses perform on par with EWC.
    \textbf{(e)}~Validation accuracy after training sequentially on subsets of Fashion MNIST with 1000 examples each. The number of training iterations was matched to joint training on the full dataset.
    Metaplastic \acp{BNN} consistently outperform networks without synaptic complexity.
    Figure adapted from \citet{laborieux_synaptic_2021}.}
    \label{fig:meta_BNN}
\end{figure}

\citet{laborieux_synaptic_2021} introduced another metaplastic model with complex synapses inspired by the Cascade model that proved useful for continual learning in \acp{BNN}  \citep{courbariaux_binarized_2016}.
A \ac{BNN} is defined as an \ac{ANN} in which synaptic weights take binary values $w_{\rm b} = \pm 1$ instead of real values (Fig.~\ref{fig:meta_BNN}a).
Although \ac{BNN} weights are binary, training them still requires a real-valued ``hidden weight'' $w_{\rm h}$ associated to each synapse, from which the binary weight is derived by taking the sign $w_{\rm b} = {\rm sign}(w_{\rm h})$.
The hidden weight is updated by gradient descent on the loss function evaluated at the binary parameters:
\begin{equation}
    \label{eq:bnn_training}
    w_{\rm h} = w_{\rm h} - \eta \nabla_{w_{\rm b}} \mathcal{L} \quad .
\end{equation}
These training dynamics turn out to be strikingly different from traditional \acp{ANN}.
For instance, the hidden weight in \acp{BNN} can grow indefinitely without convergence (Fig.~\ref{fig:meta_BNN}b).
Since the magnitudes of the hidden weights do not contribute to the network's synaptic currents, they can be viewed as an internal state of the binary synapse.
Therefore, a natural question is: What is the difference between two binary synapses with the same efficacy but different hidden weights after learning? Or more generally, how should we interpret the functional role of the hidden weight value?

\citet{laborieux_synaptic_2021} showed that the magnitude of the hidden weight naturally encodes an importance metric for the associated binarized weight, whereby higher hidden weight magnitude correspond to more important weights.
The intuition is that if the loss is evaluated at the same binary weight, the hidden weights accumulate information about the gradient at this point without changing the sign (Fig.~\ref{fig:meta_BNN}b).
Based on this observation, \cite{laborieux_synaptic_2021} designed a consolidation scheme inspired by the Cascade model \citep{fusi_cascade_2005}, in which the depressing synaptic updates are multiplied by a metaplastic factor that decreases as a function of the hidden magnitude (Fig.~\ref{fig:meta_BNN}c), thereby consolidating important synapses.
With such metaplasticity BNNs can continually learn sequences of tasks such as permuted MNIST, on par with \ac{EWC} (Fig.~\ref{fig:meta_BNN}d).
Importantly, the consolidation mechanism is agnostic to transitions between tasks, and does not rely on an additional regularization loss.
This feature prompted  the authors to explore within-task forgetting in a streaming data setting in which different parts of the dataset were available at any give time (Fig.~\ref{fig:meta_BNN}e).
When learning Fashion MNIST or CIFAR-10 in this setting, the metaplastic network outperformed the control network without complex metaplastic synapses by a large margin, close to the performance level of joint training on complete dataset. 
Thus, metaplastic BNNs show that continual learning through synaptic consolidation can be achieved locally with complex synapses and agnostic to task boundaries in deep \acp{BNN}.
It remains an exciting open question for future work to see whether and how similar principles can be extended to non-binary synapses.

\subsection{Improving memory lifetime through context-dependent gating}

So far, we have focused on purely synaptic mechanisms and how they can improve a memory system's continual learning properties. As mentioned earlier, we will forgo a discussion of replay mechanisms and system-level consolidation. These topics are discussed elsewhere in this book. However, one class of mechanisms mainly shines in combination with synaptic mechanisms. 
These are the so-called context-dependent gating mechanisms that we will discuss now.

\begin{figure}[tb]
    \centering
    \includegraphics[scale=1.0]{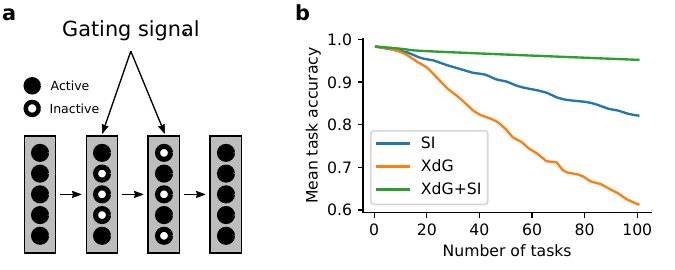}
    \caption{Context-dependent gating boosts continual learning performance of synaptic consolidation methods. \textbf{(a)}~Schematic of context-dependent gating. A task-dependent gating signal selectively turns units in the hidden layers on and off during processing of a specific task.
    \textbf{(b)}~Mean task accuracy as a function of number of tasks on the permuted MNIST benchmark for different continual learning strategies. Context dependent gating (XdG) performs poorly on its own, but outperforms other approaches when combined with \ac{SI}.
    Data re-plotted from \citet{masse_alleviating_2018}.}
    \label{fig:gating}
\end{figure}

The idea of context-dependent gating is simple. It requires a context signal that is associated with the task that is currently being learned or recalled. This signal is used to gate a subset of units in each network layer on and off. In doing so, the gating signal effectively creates a sub-network embedded within the larger network in which memories are stored (Fig.~\ref{fig:gating}). 
By allowing overlap between the sub-networks that code for different tasks, the approach still allows for forward transfer, i.e., that learned representations on previous tasks can be used in the present task, and backward transfer, i.e., that representations learned on the current task may be used by future tasks.

Surprisingly, context-dependent gating alone is not competitive compared to synaptic consolidation approaches like \ac{SI}. 
However, when combined, the two synergistically work together, improving continual learning performance \citep{masse_alleviating_2018}.
While it is not entirely clear whether or how neurobiology uses similar principles, the notion of gating can be pretty flexible.
For instance, gating does not have to take place at the single neuron level but could, in principle, also operate at the level of microcircuit, dendritic branches \citep{iyer_avoiding_2021}, or neurons and synapses \citep{barry_gateon_2023}.
In the end, it seems more than likely that neurobiology has evolved to leverage a combination of mechanisms that facilitate continual learning in the brain.

\section{Conclusion}

Throughout life, living beings must acquire new memories or skills, which require storing information in neural networks. 
These memories are long-lasting, and some of them accompany us for life. 
Still, new memories are acquired sequentially with minimal interference with previously learned knowledge. 
How do the brain's neuronal circuits achieve this feat?
The theoretical research surveyed in this chapter paints an increasingly clear picture. 
Fast storage and long memory lifetimes require memory systems with internal states and dynamics evolving over a vast range of timescales.
Plasticity algorithms can exploit such dynamics for efficient continual learning.
Presumably, our brains use this principle extensively at the synaptic, neuronal, and circuit level.
In this chapter, we have given an overview of the breadth of theories on synaptic mechanisms. 

Theories of synaptic memory consolidation are classically motivated through top-down and bottom-up considerations, and deep mechanistic insights often lie at the interface of the two. On the one hand, bottom-up models have focused on modeling experimental data on synaptic consolidation and synaptic tagging. To that end, these models had to include synaptic state and complexity to account for the experimental observations. 
Normative theories, on the other hand, have taken a top-down approach. They evolved from fundamental considerations because every memory system’s capacity is finite. As humans and animals continually learn, the finite memory capacity of their brains calls for efficient strategies that make good use of the available capacity by weaving new memories into the same synaptic fabric that already holds precious memories, ideally with minimal harmful interference between them.

Interestingly, the resulting theories from such top-down considerations also assert synaptic complexity for efficient memory storage. Thus, synaptic complexity lies at the confluence of top-down and bottom-up theoretical work. The putative role of such synaptic complexity is to address the stability-plasticity dilemma, i.e., to decide when to change their synaptic efficacy in response to current plasticity demands. 

Researchers formulated early theories of this dilemma in the \ac{IOF}, which allowed exploring the fundamental limits of the theoretical decodability of the memory trace left in the synaptic fabric. 
A central insight from these studies was that synaptic complexity can help synapses decide when to change their weight by adding synaptic complexity, corresponding to latent dynamics or an internal state at the very synapse that does not necessarily translate into a measurable change in synaptic efficacy. 
Mechanistically, individual synapses can leverage this complexity to retain a memory of how useful or necessary they are for previously stored memories and use this knowledge to modulate their propensity to undergo future change. 
Important synapses, in particular, can render themselves more resistant to future change, thereby protecting memories from premature erasure and, hence, more efficient use of the available memory capacity.

While early theories relied on an ideal observer, we need concrete network implementations and a ``actual observer'' to enjoy the benefits of added memory lifetime. 
Because instantiating plausible network models that act on real-world data is a significant feat, recent studies leveraged deep neural networks to study continual learning. 
Deep neural networks succumb to catastrophic forgetting when trained with standard gradient-based algorithms on different tasks sequentially. 
That is, they forget the earlier tasks as they acquire new tasks. We have glimpsed how simple algorithms like EWC and SI can mitigate such forgetting by leveraging synaptic complexity. 
To that end, individual synapses track their importance over past tasks to regulate future plasticity. 
While many details remain elusive, it seems likely that actual brains leverage similar metaplasticity mechanisms to learn continually \citep{jedlicka_contributions_2022}.

While this is exciting progress in our conceptual understanding of the principles enabling neural networks to learn continually, fundamental questions remain open. On the one hand, future work must part with the rigorous shackles of task-based incremental learning and move toward task-free approaches \citep{van_de_ven_three_2022}. 
In most, but not all, of the work discussed above, the identity of the task and when it ends are known. This task-awareness contrasts many real-life situations in which transitions between one and the other are fluid, and a task label may not be available. Consider, for instance, driving through a mountainous landscape and gradually transitioning into a desert landscape.
It will be essential for future research to dive into the question of how actual brains cope with this challenge.

Another issue that arises with many current continual learning paradigms is the phenomenon of loss of plasticity. While many different scenarios can lead to loss of plasticity, one that intelligent synaptic learning algorithms can succumb to is the one in which weights get increasingly consolidated, eventually preventing any future learning. Our brains do not suffer from such a loss of plasticity. Although we tend to learn slower as age progresses, we hardly lose our ability to learn anything new. While some exciting ideas and models try to tackle the issue of loss of plasticity, it will likely remain an important research question for both \ac{AI} and neuroscience.

Finally, we must progress in connecting our top-down motivated models that have primarily been derived using a deep learning framework to phenomenological plasticity models and realistic spiking neural network models. This step will allow interfacing theory and experiment and is thus essential to test whether and how neurobiology leverages similar continual learning principles. 
In particular, we have to work towards integrative theories that combine insights from neurobiology, such as cross-tagging, consolidation, and reconsolidation \citep{alberini_mechanisms_2005, gershman_computational_2017}, which presently do not have an apparent correspondence in the functional continual learning models developed for deep neural networks. 
Studying these phenomena may also shed light on why humans and animals are typically better at blocked training \citep{flesch_comparing_2018}, whereas \acp{ANN} still excel at joint training. 
To address these questions, we must derive testable predictions from the models and systematically verify them through experiments. 
Doing so will likely require a multi-scale approach in which we derive simultaneous predictions at the network, microcircuit, neuronal, and synaptic levels. 
Models that accurately capture continual learning in the brain must also accurately account for effects across scales.

In summary, combining ideas from theoretical neuroscience and deep learning has led to fresh ideas of how our brains might deal with the plasticity-stability dilemma. 
This synergy has already created a wealth of new insights that add to a rich repertoire of phenomenological models. 
Still, we are only beginning to scrape the surface of understanding continual learning in the brain. 
We should further deepen this understanding by carefully aligning top-down and bottom-up theories. 
Additionally, we need tangible model predictions and the associated targeted experiments for iteratively validating the theories. 
Exploring these ideas  in coming years will be exciting.
Still, in the long run, it requires continuous efforts in training young scientists who are fluently versed in ideas from machine learning as well as theoretical and experimental neuroscience.

\section*{Acknowledgements}

This work was supported by the Swiss National Science Foundation [grant numbers PCEFP3\_202981 and TMPFP3\_210282] and the Novartis Research Foundation.

\bibliographystyle{elsarticle-harv}

\end{document}